\documentclass[twocolumn,12pt]{iopart}

\usepackage{graphicx}
\usepackage{dcolumn}
\usepackage{color}

\begin{document}

\title[]{Control of transverse motion and X-ray emission of electrons accelerated in laser-driven wakefields by tuning laser spatial chirp}

\author{H Xiao$^{1,2}$, M Chen$^{1,2,*}$, S M Weng$^{1,2}$, L M Chen$^{3,2}$, Z M Sheng$^{1,2,4}$, J Zhang$^{1,2}$}
\address{$^1$Key Laboratory for Laser Plasmas (MoE) and School of Physics and Astronomy, Shanghai Jiao Tong University, Shanghai, 200240, China}
\address{$^2$Collaborative Innovation Center of IFSA (CICIFSA), Shanghai Jiao Tong University, Shanghai 200240, China}
\address{$^3$Beijing National Laboratory of Condensed Matter Physics, Institute of Physics, CAS, Beijing 100190, China}
\address{$^4$SUPA, Department of Physics, University of Strathclyde, Glasgow G4 0NG, UK}
\ead{*minchen@sjtu.edu.cn}
\vspace{10pt}

\begin{abstract}
A method is proposed to control the transverse motion and X-ray emission of an electron beam in a laser driven wakefield by tuning the laser spatial chirp. The dispersion of a transversely chirped laser pulse and the transverse nonuniform refractive index of the plasma channel result in transverse laser centroid oscillations, which leads to periodic transverse oscillations of the laser-driven wake. Electrons accelerated inside the wake also undergo transverse oscillations making wiggler like motion. Both the oscillation period and amplitude can be controlled by tuning the laser chirp, the pulse duration or the plasma channel width. As a result, the far field spectral distributions of the X-ray emission can be flexibly manipulated.

\end{abstract}

\maketitle
\ioptwocol

\section{Introduction}

Compact x-ray radiation generation is a promising application of laser wakefield acceleration due to the ultra-high accelerating gradient of the wake \cite{Esarey2009,Gonsalves2019} and the improvement of the beam quality in recent years \cite{Wang2016,Mirzaie2015,Zeng2015,Osterhoff2008,Rechatin2009}. Radiation can be generated from the electron betatron oscillations inside the wake or from the interaction of the electron beam with external magnetic or optical wigglers \cite{Corde2013,Hooker2013,Kneip2010,Schlenvoigt2008,Fuchs2009,Phuoc2012,Yi2016,Huang2016,Yan2017,Wenz2019}.The former is easy for operation but with inflexible tunability both on spectrum and intensity since the oscillation period $\lambda_\beta=2\pi/k_\beta$ and amplitude $r_\beta$ are varying along the beam propagation \cite{Corde2013,Esarey2002}. The latter is much flexible for radiation tunability, however, the coupling between the electron beam and the wiggler is difficult, and the device is relatively large and costs too much for laboratories in universities and small institutes. Improved schemes were recently investigated by using a plasma channel to induce controllable transverse oscillation of the laser pulse and the wakefield behind the laser driver \cite{Lee2015,Rykovanov2015,Chen2016,Luo2016,Rykovanov2016}. In these plasma-based schemes, the usual downstream undulator/wigger part is naturally integrated into the plasma accelerator. Electrons undergo simultaneously longitudinal acceleration and transverse oscillation inside the channel.

Depending on the different ratios between the laser centroid oscillation period ($\Lambda_{o}$) and the electron betatron oscillation period inside the wake ($\lambda_\beta$), different radiation enhancement occurs. Lee \textit{et al.} studied the case where $\lambda_\beta\ll\Lambda_o$ and found that in this case the whole wake oscillation can enhance the oscillation amplitude of the  betatron radiation, which leads to higher critical energy of emitted photons \cite{Lee2015}. Rykovanov \textit{et al.} studied the case where $\lambda_\beta\gg\Lambda_o$ and found that in that case the whole wake oscillation can be used as a plasma undulator for fundamental and high order harmonics radiation generation by injecting an electron beam at the point with zero longitudinal wake field \cite{Rykovanov2015}. Lei \textit{et al.} studied the case where $\lambda_\beta\approx\Lambda_o$ and found that the oscillation amplitude can be significantly enhanced when the electron betatron oscillations are in resonance with the laser centroid oscillations, which may extend the radiation spectrum to $\gamma$-ray range \cite{Lei2018}. Chen \textit{et al.} studied the wiggler like synchrotron radiation from the ionization injected electrons in a case with stronger strength parameter \cite{Chen2016}. And recently Luo \textit{et al.} have investigated this in a more general case with three dimensional particle-in-cell (PIC) simulations where they found a helical undulator like radiation with controllable polarization can be generated \cite{Luo2016}.

The transverse oscillation of an off-axis injected laser pulse in a plasma channel is due to the refractive index gradient depending on the transverse plasma density profile. Here we consider a parabolic plasma channel with electron density distribution $n(r)=n_0+\Delta nr^2/r_0^2$, where $\Delta n$ is the channel depth and $r_0$ is the channel width. The refractive index is a function of the drive laser pulse frequency as $\eta_r=1-[1+({\Delta n }/{n_0})({ r^2}/{r_0^2})]{\omega_{p0}^2}/{2\omega^2} $, where $\omega_{p0}$ represents the on-axis plasma frequency and $\omega$ is the frequency of the drive laser pulse.  The refractive index is determined by the transverse plasma density distribution, and, if the laser has a transverse spatial chirp, i.e., $\omega=\omega(r)$, there is an additional transverse dependence. The radial dependence of the index of refraction $\eta_r (r)$ will cause the laser evolution and centroid motion. This paper studies the laser transverse chirp effects on the laser propagation in a plasma channel and the following laser-driven wakefield. The transverse motion and betatron radiation of the electron beams accelerated from such wakefields are investigated as well. We find tunable radiation can be obtained by laser spatial chirp tuning.

\section{Transverse laser chirp effects on pulse evolution and centroid oscillation}

\begin{figure}
\includegraphics[width=8.2cm]{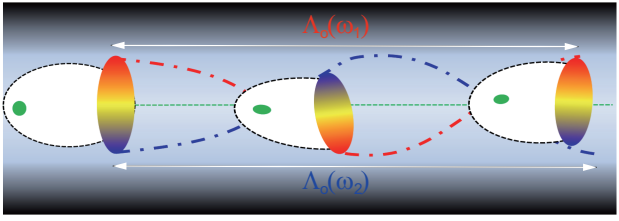}
\caption{Schematic of laser transverse-chirp-controlled wakefield acceleration and beam transverse motion. The oscillation periods of the different color components of the laser pulse are frequency dependent, leading to pulse tilting, transverse chirp reversal, and wakefield oscillations.}\label{schematic}
\end{figure}

A schematic view of laser pulse evolution and centroid motion is shown in Fig.~\ref{schematic}. A laser pulse with normalized electric field of $a=eE/(m\omega_0 c) \simeq a_0 \exp(-r^2/W_0^2)\exp(-\xi^2/L_0^2)\rm {cos}[(\omega_0+\omega')t]$ is injected on-axis into a plasma channel, where $r=\sqrt{y^2+z^2}$ represents the transverse distance from the axis, $\omega_0$ is the central frequency of the laser and $\omega'$ is a function of space and describes the spatial chirp, $\xi=x-v_gt$, and $v_g$ is the group velocity of the pulse. An unmatched laser pulse propagating inside a plasma channel with a parabolic density profile will experience periodic self-focusing and defocusing. The periodic length is $\lambda_u=\pi Z_M(\Delta n_c/\Delta n)^{1/2}$ with $Z_M=\pi r_0^2/\lambda_0$ and $\Delta n_c=1/\pi r_e r_0^2=1.13\times10^{20}(cm^{-3})/r_0^2(\mu m)$, where $r_e=m_ec^2/e$ is the classical electron radius \cite{Esarey2009}. Since $\lambda_u$ is laser wavelength dependent, the light rays initially at different transverse positions have different color and they will follow different trajectories. For example, the red and blue dashed lines in Fig.~\ref{schematic} show two typical rays of the propagation when the laser pulse has a transverse linear chirp, i.e., $\omega'=\alpha y\omega_0/W_0$, where $\alpha$ is a constant representing the spatial chirp degree in the form of normalized frequency gradient. In addition to the trajectory differences, the laser phase velocity is also different, i.e., $v_p=c [1-\omega_p^2/\omega^2]^{-1/2} \simeq c(1+\omega_p^2/2\omega^2)$. Due to these two effects, the pulse will experience transverse deformation and pulse front tilting, which then leads to the laser centroid oscillations. The laser centroid oscillation period satisfies $\Lambda_o\simeq2\lambda_u$ \cite{Rykovanov2015,Chen2016,Rykovanov2016,Liu2010}.

\begin{figure}
\includegraphics[width=8.2cm]{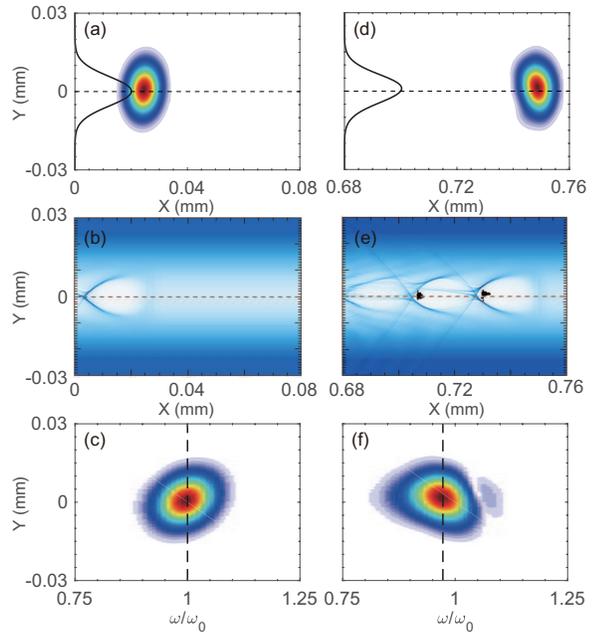}
\caption{(a,d) Spatial distributions of the laser pulse intensity profile $E_z^2$, (b,e) the plasma density of the wake and electrons trapped in the first two bubbles, and (c,f) the laser spectral distribution along the transverse direction $\tilde{E}(\omega, y)= \vert \int E(x,y)\exp(i\omega x/c) dx \vert$. Figs.(a,b,c) correspond to the laser just entering into the plasma channel and Figs.(d,e,f) correspond to the laser after propagating in the channel for 0.75~mm.}\label{laserwakespectrum}
\end{figure}

To investigate these effects on wakefield acceleration, especially the controlled transverse dynamics of the electrons, we use the VLPL particle-in-cell code \cite{Pukhov1999} to model the interaction. In the simulations, a linearly polarized laser pulse with electric field along the $z$ direction and normalized electric field intensity of $a_0=2.0$ enters into the plasma channel at $t=0$ from the left boundary of the simulation box. The focused radius of the pulse is $W_0=10\lambda_0$, where the laser central wavelength is $\lambda_0=0.8\mu m$. The on-axis density of the plasma channel is $n_0=0.001n_c$, and $\Delta n=\Delta n_c$. The simulation box has dimensions of $80\times60\mu m^2$ and is divided into $2400\times600$ cells, along $x$ and $y$ directions, respectively. To observe the laser chirp, the pulse duration and channel width effects on the wakefield acceleration and the beam transverse oscillation, we vary the linear transverse chirp of the pulse ($\alpha$), the duration of the pulse ($L_0$) and the radius of the channel ($r_0$) independently.

Figure~\ref{laserwakespectrum} shows a typical simulation with laser transverse chirp of $\alpha=0.0125$, the pulse duration of $L_0=6T_0$, where $T_0=2\pi/\omega_0$ and plasma channel width of $r_0=W_0$. The initial laser intensity profile and wakefield are shown in Figs.~\ref{laserwakespectrum}(a) and (b), respectively. The laser frequency distribution along $y$ direction is shown in Fig.~\ref{laserwakespectrum}(c), obtained by performing a Fourier transformation of the laser electric field along the longitudinal direction ($x$). The dashed black line represents the central frequency.  As one can see the pulse initially has a higher frequency component in the top half plane ($y > 0$). The laser and wakefield after propagating about 0.75 mm are shown in Fig.~\ref{laserwakespectrum}(d,e,f). One can see that the transverse centroid of the laser pulse has moved to the top half plane( $y > 0$), and both the wake and the accelerated electrons inside the buckets have deviated from the channel axis. The transverse laser chirp has rotated its direction (sign), as shown in Fig.~\ref{laserwakespectrum}(c,f), and the central laser frequency has decreased due to the wake excitation, as shown by the black dashed lines. This laser evolution is consistent with the predictions, schematically shown in Fig.~\ref{schematic}. The simulation of the laser propagation up to 5 mm exhibits periodic laser pulse oscillation and spectrum rotation.

\begin{figure}
\includegraphics[width=8.2cm]{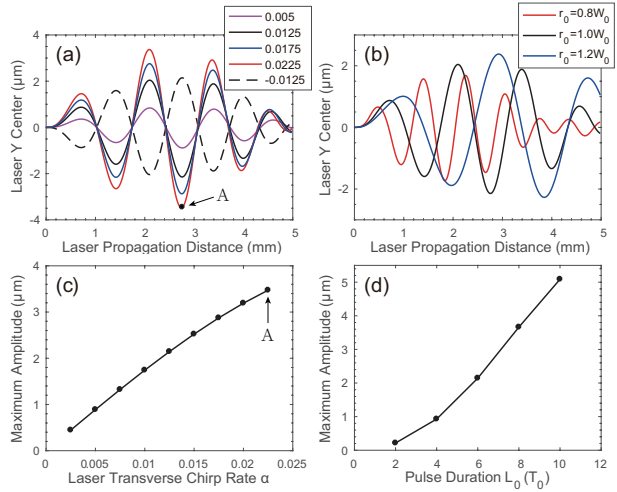}
\caption{(a) Evolution of laser centroid oscillation along the propagation direction for different transverse laser chirps, with the plasma channel width fixed at $r_0=W_0$ and the pulse duration is fixed at $L_0=6T_0$. (b) Evolution of laser centroid oscillation for different plasma channel widths, with the transverse laser chirp of $\alpha=0.0125$ and the pulse duration of $L_0=6T_0$. (c,d) shows the evolution of the absolute value of the second negative peak of the laser centroid oscillation along with the laser chirp rate $\alpha$ and pulse duration $L_0$, respectively.}\label{lasercentroid}
\end{figure}

The laser transverse oscillation is tunable when the transverse laser chirp, the pulse duration or the plasma channel radius are varied. In Fig.~\ref{lasercentroid}(a), typical transverse laser centroid oscillations resulting from different initial transverse chirps are shown. The initial increasing of the centroid oscillation amplitude during laser propagation, indicating that the optimal chirp profile for maximum oscillation may not be a linear profile. For our linear chirp cases, serial simulations show that the maximum oscillation amplitude is proportional to the spatial chirp rate $\partial \omega'/\partial y=\alpha \omega_0/W_0$ when $\alpha$ is small, as shown in Fig.~\ref{lasercentroid}(c). Besides the linear chirp rate, increasing the pulse duration can also enhance the transverse oscillation, which is shown in Fig.~\ref{lasercentroid}(d). These studies mean that the beam transverse oscillation amplitude is easily tunable through tuning laser pulse parameters. We should mention that usually the focused width of the laser pulse would vary with the linear spatial chirp as {$W{'}/W_0=(1-\alpha^2 \omega_0^2 \tau_0^2 /4)^{-1/2}$} with $W{'}$ represents the overall beam width increased from $W_0$ due to spatial chirp and $\tau_0$ is the full width at half maximum (FWHM) of the pluse duration \cite{Akturk2004}. In our simulation parameters, $W{'}/W_0\le1.412$ which includes little influence in final result. Too large chirp rate will lead to a non-Gaussian pulse, which affects the wake excitation and acceleration. It is out of our interests.

The transverse oscillation period ($\Lambda_o$) does not depend on the laser pulse. It is determined by the channel structure as $\Lambda_o$ states. The calculated linear period length is approximately $\Lambda_o=1.579$~mm, and the simulation results show the length of the 3 periods over the 5 mm acceleration distance in Fig.~\ref{lasercentroid}(a) are 1.743, 1.342 and 1.239 mm. They are close to the calculated value. The difference is due to the nonlinear effects for $a_0=2$ and the laser frequency downshift during the propagation. The period depends on the channel width as $\Lambda_o \propto r^2_0$, and the dependence on $r_0$ is shown in Fig.~\ref{lasercentroid}(b). A wider channel with $r_0=1.2W_0$ gives a longer oscillation period. The linear value of the wider channel is $\Lambda_o=2.274$~mm and the two period lengths of the blue solid line in Fig.~\ref{lasercentroid}(b) are 2.420 and 1.906 mm. For the case of $r_0=0.8W_0$, the calculated period length is $\Lambda_o=1.011$~mm, and the first two period lengths of the red solid line in Fig.~\ref{lasercentroid}(b) are 1.165 and 0.880 mm. One can also see from Fig.~\ref{lasercentroid}(a) that when the sign of the initial chirp has been changed, from $\alpha=0.0125$ to $\alpha=-0.0125$, the initial oscillation direction of the laser centroid changes. After a few periods of propagation, the central laser frequency downshifts, owing to laser energy deposition into the plasma wave, and the spatial chirp is reduced with a reduction in the laser centroid oscillation as well. A more stable oscillation usually needs a longer driver pulse.

\section{Transverse laser chirp effects on electron beam oscillation and betatron radiation}

\begin{figure}
\includegraphics[width=8.2cm]{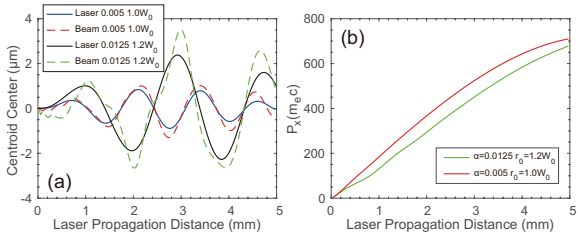}
\caption{(a) Centroid oscillation of the laser and electron beam for two sets of laser transverse chirp and plasma channel width. (b) Electron beam energy evolution along the laser propagation for two sets of laser transverse chirp and plasma channel width.}\label{beamcentroid}
\end{figure}

Provided the laser oscillation amplitude is not too large, one would expect that the electrons in the wake behind the driver pulse will oscillate following the laser centroid oscillation. To see the relationship between these two kinds of oscillations  Fig.~\ref{beamcentroid}(a) shows both of the laser and electron beam centroid oscillations for two sets of simulations with different laser transverse chirps and plasma channel widths. The beam has been injected into the wakefield using ionization injection \cite{Chen2006, Pak10,Chen2012}. A short stage of Nitrogen gas with density of 0.0005 $n_c$ and length of 20 $\lambda_0$ is inserted into the beginning of the plasma channel from $x=30\lambda_0$ to $x=50\lambda_0$ with an up-ramp-plateau-down-ramp ($5\lambda_0-10\lambda_0-5\lambda_0$) profile.

In Fig.~\ref{beamcentroid}(a) the solid lines show the laser centroid oscillation and the dashed lines show the corresponding electron beam oscillation. One can see when the laser oscillation amplitude is small (shown by the solid blue line), the electron beam oscillation (shown by the dashed red line) tracks the laser oscillation. The electron beam oscillation includes both the laser centroid oscillation, with period of $\Lambda_o$, and the betatron oscillation, with period of $\lambda_\beta=\sqrt{2\gamma}\lambda_p$ \cite{Rykovanov2016}. The latter depends on the energy of the electron beam. When the laser oscillation amplitude increases (shown by the solid black line), the deviation between the electron beam centroid (shown by dashed green line) and the laser centroid increases. Loss of electrons will occur for sufficiently large oscillation amplitude, which usually happens when the transverse spatial chirp is too large. Within our study parameters ($\alpha<0.0225$), no electron transverse loss has been observed. As one can also see from Fig.~\ref{beamcentroid}(b) that larger amplitude oscillations lead to smaller the average acceleration gradient. This would also affect the final radiation spectrum.

\begin{figure}
\includegraphics[width=8.2cm]{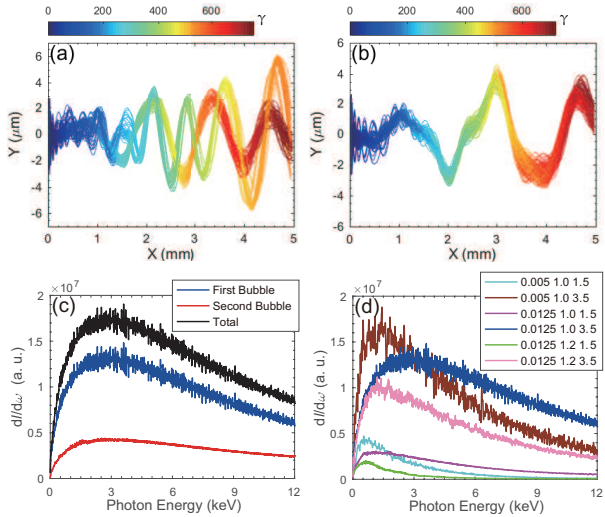}
\caption{(a) Typical trajectories of the accelerated electrons in the first and second buckets. The width of plasma channel here is $r_0=W_0$. (b) Typical trajectories of the accelerated electrons in the first bucket using the same simulation parameter with (a) except for $r_0=1.2W_0$. (c) The on-axis radiation spectra from electrons accelerated in the first (blue line), second buckets (red line), and the total spectrum (black line). (d) Radiation spectra from the accelerated electrons with different sets of laser transverse chirp, acceleration distance and width of the plasma channel. The legend in (d) shows the value of $\alpha$, $r_0/W_0$ and $L_{\rm acce}$ (mm).}\label{radiationspectrum}
\end{figure}

To study the radiation properties of these transversely oscillated electrons, we calculate the radiation spectrum by using a post-processing code VDSR \cite{Chen2013}. This code numerically simulates the final spectrum in a way of incoherent radiation addition of the selected electrons whose trajectories are imported from the PIC simulations. We randomly choose 100 electrons trapped by the wake to concentrate on investigating the collective radiation properties due to the limitation of computational resource. The real radiation intensity is proportional to the charge of the injected electrons but the shape of the radiation spectrum does not vary too much, which is demonstrated in a special calculating case where the particle number used for the radiation calculation is doubled. Typical trajectories of randomly selected electrons are shown in Fig.~\ref{radiationspectrum}(a,b), which correspond to the simulation parameters of $r_0=W_0$ and $r_0=1.2W_0$, respectively, with the same laser transverse chirp $\alpha=0.0125$. In Fig.~\ref{radiationspectrum}(a), one can clearly see that actually there are two sets of electrons. They come from the first two acceleration bubbles as shown in Fig.~\ref{laserwakespectrum}(e). In the simulation we found that the ratio of the acceleration charge and average gamma factor of the electrons in the first and second bubbles are about 2.5 and 1.3, respectively. The radiation spectra of electrons from the first bubble, the second bubble and both of them are shown in Fig 5(c). Less charge and lower energy of the accelerated electrons in second bubble make their radiation much weaker and tend to lower radiation energy, so in the following radiation tuning studies we only consider the radiation from electrons trapped in the first acceleration bubble.

To avoid the waste of the calculation time on the irregular oscillations and the correspondingly low energy radiation of the accelerated electrons at the beginning of acceleration, we cut off the first $1.5\rm mm$ long trajectories when we calculate the final radiation spectrum. Radiation spectra from different sets of plasma channel width ($r_0$), laser transverse chirp parameter ($\alpha$) and acceleration distance ($L_{\rm acce}$) are shown in Fig.~\ref{radiationspectrum}(d). As one can see, both the peak positions and amplitudes of the radiation spectra vary with these parameters. A longer acceleration distance means higher electron energy, which makes the radiation tend to high frequency part and the radiation intensity increase. As we show before, increasing the laser transverse chirp leads to higher transverse oscillation amplitude, which also increases the radiation intensity. A wider channel means a larger trajectory period and corresponds to a lower radiation frequency. All these radiation characters depending on the laser and channel properties are demonstrated in Fig.~\ref{radiationspectrum}(d). It means in a larger range of photon energy (1-10keV), the spectrum can be flexibly tuned.

\section{Summary and discussion}

In summary, the control of transverse motion and X-ray emission of electrons accelerated in laser-driven wakefields can be realized by using laser spatial chirp tuning. We noticed that the laser spatial chirp effects on wakefield acceleration in a uniform plasma has been studied experimentally by Popp \textit{et al.} \cite{Popp2010}, who found that by tuning the alignment of the grating pair in the laser compressor, the transverse chirp can be tuned. The resulting laser pulse will have a pulse-front tilt during the propagation that can provide electron-beam steering. The introduction of plasma channel in this work enables beam oscillation control.  In particular, the effect of laser spatial chirp on laser propagation in a plasma channel has been studied, as well as the effect on beam acceleration in the laser-driven wakefield. Due to the dispersion of the laser pulse along the transverse direction, its different frequency components experience different oscillation trajectories, which causes the laser centroid and the trailing wake to carry out transverse oscillations. The oscillation period can be controlled by tuning the plasma channel parameters and the oscillation amplitude can be tuned by varying the transverse laser chirp. The accelerated beam inside the wake performs similar transverse motion. The mechanism provides a method for controlled x-ray radiation from these oscillating electron beams. Both the peak position and amplitude of the radiation spectrum can be controlled through tuning the above two parameters.

\ack
This work is supported in part by the Science Challenge Project(No.TZ2018005), the National Science Foundation of China (11774227). M.C. thanks C.B. Schroeder and K. Nakamura at Lawrence Berkeley National Laboratory for their helpful discussions on the laser transverse chirp effects. Simulations were performed on the $\rm \Pi$ supercomputer at Shanghai Jiao Tong University.

\end{document}